\input harvmac 
\input epsf.tex 

\lref\lmr{Simone Lelli, Michele Maggiore and Anna Rissone,
``Perturbative and nonperturbative aspects of the
two-dimensional string / Yang-Mills correspondence," Nucl.Phys. 
{\bf B656} 37-62, (2003), {\tt hep-th/0211054}.}
\lref\dewiti{Gabriel Lopes Cardoso, Bernard de Wit and
 Thomas Mohaupt, ``Macrosopic
entropy formulae and nonholomorphic
 corrections for supersymmetric black holes,"
Nucl. Phys. {\bf B567} 87-110, (2000), {\tt hep-th/9906094}.} 
\lref\vaftdym{Cumrun Vafa, ``Two Dimensional 
Yang-Mills, Black Holes and Topological Strings," {\tt 
hep-th/0406058}.} 
\lref\mig{A.~Migdal, ``Recursion Equations In Gauge Field Theories,"
{\it Zh. Eksp. teor. Fiz. } {\bf 69}, 810 (1975).} 
\lref\grotayi{D.~J.~Gross and W.~I.~Taylor,
``Two-dimensional QCD is a string theory,''
Nucl.\ Phys. {\bf B400}, 181 (1993) {\tt hep-th/9301068}.}
\lref\grotayii{David J. Gross and Washington Taylor,
 ``Twists and Wilson Loops in the
String Theory of two-dimensional QCD," Nucl. Phys. 
{\bf B403} 395-452,(1993), {\tt hep-th/9303046}.} 
\lref\doug{ M.~R.~Douglas, ``Conformal Field Theory Techniques for Large N Group Theory," {\tt hep-th/9303159}.}
\lref\jev{A. Jevicki, ``Nonperturbative Collective Field Theory," Nucl.Phys. {\bf B376} 75-98, (1992).}
\lref\minpoly{Joseph A. Minahan and Alexios P. Polychronakos, ``Equivalence of 
two-dimensional QCD and the c=1 matrix model," Phys.Lett. {\bf B312} 155-165, (1993), {\tt hep-th/9303153}.}
\lref\cmr{Stefan Cordes, Gregory W. Moore and Sanjaye Ramgoolam,
 ``Lectures on 2-D 
Yang-Mills theory, Equivariant Cohomology and Topological Field
 Theories," Nucl. Phys.
Proc. Suppl. {\bf 41} 184-244, (1995) {\tt hep-th/9411210}.} 
\lref\hor{Petr Horava, ``Topological Strings and QCD in
 two-dimensions," {\tt hep-th/9311156}.}
\lref\osv{Hirosi Ooguri, Andrew Strominger,  and Cumrun Vafa, 
``Black Hole Attractors and 
the Topological String," Phys. Rev. {\bf D70}:106007, (2004), 
{\tt hep-th/0405146}.} 
\lref\vaf{ C.~Vafa, ``Two Dimensional Yang-Mills, Black Holes and Topological Strings,"
{\tt hep-th/0406058}. } 
 \lref\Witten{  E.~Witten, ``D-Branes And K-Theory,"
JHEP {\bf 9812}, 019 (1998)  {\tt hep-th/9810188}.}
\lref\gss{ L. Griguolo, D. Seminara, R.J. Szabo, ``Double Scaling String Theory of QCD in Two Dimensions," {\tt hep-th/0409288}.}
\lref\ram{S. Ramgoolam, ``On spherical harmonics for fuzzy spheres in diverse dimensions," Nucl.Phys. {\bf B610} (2001) 461-488
{\tt hep-th/0105006}.} 
\lref\clt{J.~Castelino, S.~Lee and W.~Taylor, ``Longitudinal 5-branes as 4-spheres in 
Matrix theory," Nucl.Phys. {\bf B526} (1998) 
334-350, {\tt hep-th/9712105}.}
\lref\nabint{Neil R. Constable, Robert C. Myers, Oyvind Tafjord, ``Non-abelian Brane
Intersections," JHEP {\bf 0106} (2001) 023, {\tt hep-th/0102080}.} 
\lref\nara{V.P. Nair, S. Randjbar-Daemi, ``On brane solutions in M(atrix) theory,"
Nucl.Phys. {\bf B533} (1998) 333-347, {\tt hep-th/9802187}.} 
\lref\grostei{Harald Grosse, Harold Steinacker, ``Finite Gauge Theory on Fuzzy $CP^2$," Nucl.Phys. {\bf B707} (2005) 145-198, {\tt hep-th/0407089}.} 
\lref\expoas{M.~V.~Berry and C.~J.~Howls, Proc. R. Soc. Lond. {\bf A430} (1990) 653-667;
Proc. R. Soc. Lond. {\bf A434} (1991) 657-675\semi
E. Delabaere, H. Dillinger and F. Pham, J. Math. Phys. {\bf 38} (1997) 6126-6184\semi
B. Sternin and V. Shatalov, ``Borel-Laplace Transform and Asymptotic Theory," (CRC
Press, Boca Raton, FL, 1996).}
\lref\moore{Gregory W. Moore, ``K Theory from a
 physical perspective," {\tt hep-th/0304018}.}
\lref\aabf{Ofer Aharony, Yaron E. Antebi, Micha Berkooz, 
and Ram Fishman, ``Holey Sheets - Pfaffians and Subdeterminants as 
D-brane Operators in Large N Gauge Theories," 
JHEP {\bf 0212} (2002) 069, {\tt hep-th/0211152}.}
\lref\llm{Hai Lin, Oleg Lunin and Juan Maldacena,
``Bubbling AdS space and 1/2 BPS geometries,"
JHEP {\bf 0410} (2004) 025, {\tt hep-th/0409174}.}
\lref\ber{David Berenstein, ``A matrix model for a quantum hall 
droplet with manifest particle-hole symmetry," {\tt hep-th/0409115}.} 
\lref\kg{Robert de Mello Koch and Rhiannon Gwyn, 
``Giant Graviton Correlators from Dual SU(N) super Yang-Mills
Theory," JHEP {\bf 0411} (2004) 081, {\tt hep-th/0410236}.} 
\lref\cjr{Steve Corley, Antal Jevicki and Sanjaye Ramgoolam,
``Exact Correlators of Giant Gravitons from dual N=4 SYM,"
Adv. Theor. Math. Phys. 
{\bf 5} (2002) 809-839, {\tt hep-th/0111222}.}
\lref\bbns{Vijay Balasubramanian, Micha Berkooz, Asad Naqvi, and Matthew J. Strassler, ``Giant Gravitons in Conformal Field Theory," JHEP {\bf 0204} (2002) 034,
{\tt hep-th/0107119}.} 
\lref\ddmp{Atish Dabholkar, Frederik Denef, Gregory W. Moore and Boris Pioline,  ``Exact and Asymptotic Degeneracies of Small Black Holes," {\tt hep-th/0502157}.} 
\lref\sen{Ashoke Sen, ``Black Holes and the Spectrum of Half-BPS States in N=4
Supersymmetric String Theory," {\tt  hep-th/0504005}.} 
\lref\myersdiel{   R.~C.~Myers,
  ``Dielectric-branes,''
  JHEP {\bf 9912}, 022 (1999)
  [arXiv:hep-th/9910053].
}
\lref\trivai{
  S.~P.~Trivedi and S.~Vaidya,
  JHEP {\bf 0009}, 041 (2000)
  [arXiv:hep-th/0007011].
}

\lref\bata{
  J.~Baez and W.~Taylor,
  Nucl.\ Phys.\ B {\bf 426}, 53 (1994)
  [arXiv:hep-th/9401041].
}

\lref\matmat{
  T.~Matsuo and S.~Matsuura,
  Mod.\ Phys.\ Lett.\ A {\bf 20}, 29 (2005)
  [arXiv:hep-th/0404204].
}

\lref\ferr{
  F.~Ferrari,
  ``The large N limit of N = 2 super Yang-Mills, fractional instantons and
  infrared divergences,''
  Nucl.\ Phys.\ B {\bf 612}, 151 (2001)
  [arXiv:hep-th/0106192].
  }

\lref\maldasuss{
  J.~M.~Maldacena and L.~Susskind,
  Nucl.\ Phys.\ B {\bf 475}, 679 (1996)
  [arXiv:hep-th/9604042].
}

\lref\brodie{
  J.~H.~Brodie,
  Nucl.\ Phys.\ B {\bf 532}, 137 (1998)
  [arXiv:hep-th/9803140].
}

\def\YD{ Young Diagram  } 

\vskip 1cm

 \Title{ \vbox{\baselineskip12pt\hbox{  QMUL-PH-05-07  }
               \baselineskip12pt\hbox{  Wits-CTP-021    }
               \baselineskip12pt\hbox{  Brown-Het-1444   }}
}
 {\vbox{
\centerline{ On  Exponential corrections to the $1/N$ expansion in   }
\centerline{   two-dimensional Yang Mills  }  }}

\centerline{ { Robert de Mello Koch $^{1}$, Antal Jevicki $^{2}$,  
Sanjaye Ramgoolam $^{3}$  }}

\vskip.1in 
\centerline{{\sl ${}^{1}$ University of Witwatersrand  }}
\centerline{{\sl Wits, 2050  }}
\centerline{{\sl South Africa  }}
\vskip.1in 
\centerline{{\sl ${}^{2}$ Brown University }}
\centerline{{\sl Providence, RI02912 , USA  }}
\vskip.1in 
\centerline{{\sl ${}^{3}$ Queen Mary College }}
\centerline{{\sl London, E1 4NS, UK  }}
\smallskip
\centerline{{\tt robert@neo.phys.wits.ac.za, antal@het.brown.edu, 
 s.ramgoolam@qmul.ac.uk }}
\vskip .2in

We compute $e^{-AN}$ corrections to the Gross-Taylor $1/N$ expansion 
of the paritition function of two-dimensional
$SU(N)$ and $U(N)$ Yang Mills theory. 
We find a very similar structure of mixing between holomorphic 
and anti-holomorphic sectors as that
 described by Vafa for the $1/N$ expansion. 
Some of the non-perturbative terms are suggestive 
of D-strings wrapping the $T^2$ of the 2dYM but blowing 
up into a fuzzy geometry by the Myers 
effect in the directions transverse to the $T^2$.

\Date{ April   2005 }

\newsec{ Introduction   } 

The exact partition function of 2dYM on a torus $T^2$  is known and 
expressed simply as a sum over representations of $SU(N)$ \mig. 
\eqn\exctans{ 
Z = \sum_{ Y } e^{-{ A g_{YM}^2 \over2 }  C_2 ( Y ) } 
} 
Gross-Taylor \refs{ \grotayi , \grotayii } 
have described the perturbative $ 1/N $ 
expansion of the 2dYM partition function  and interpreted it in terms 
of maps of elementary string worldheets to the $T^2$.  
There is a free fermion description of this system \refs{ \jev , \doug }.  
A topological string theory was given in 
\refs{ \cmr , \hor }. Non-perturbative aspects were studied in \jev\ 
and subsequently \lmr . A different approach to non-perturbative aspects 
was pursued in \refs{ \matmat , \bata }.  
Recent work of \refs{ \osv\vaf } has renewed interest in the 
structure of the non-perturbative expansion.

The Gross-Taylor approach first constructs 
a ``chiral'' perturbative $ 1/N$ expansion 
\eqn\chirpt{ 
Z^{pert}_{chir} = \sum_{n=0}^{\infty} \sum_{ R  } e^{- A C_2( R  )   } . 
} 
Here the Casimirs of $R$ are computed under the assumption 
that $ R $ are  Young diagrams with $n$ boxes, where 
$n$ is small compared to $N$. 
If small Young diagrams contribute to the perturbative 
expansion so do their conjugates
 which have columns of length close to $N$, as well as 
 composites which have a few long columns and a few short 
columns. Taking into account these composite Young Diagrams 
gives the  complete $1/N$ expansion which is the non-chiral 
partition function
\eqn\nchirpt{ 
Z^{pert}_{nchir} = \sum_{n=0}^{\infty} \sum_{\tilde n = 0}^{\infty } 
                   \sum_{ R ,  S }   e^{- A C_2( R \bar S  )}
}

We would like to generalize these calculations to include 
Young Diagrams which contribute at order $e^{-A N } $. 
One class of such diagrams is given in \lmr.  
They consider Young diagrams containing  $ k $ rows  of length $ N $. 
$ k$ is of order $1$ as N goes to infinity.  
Attached to this rectangular block of $kN $ boxes, 
there are two small Young diagrams $R_1$ and $R_2$, with number of 
boxes of order $1$ as $N$ goes to infinity.  $R_1$ is attached 
at the bottom left and $R_2$ is attached at the right. 
As observed in \lmr\ we also need to  consider,
 along the lines of \refs{\grotayi, \grotayii }, 
 the $SU(N)$ duals of these Young diagrams, 
and subsequently form the chiral-antichiral composites.

There is a simple generalization of this kind of Young diagram 
which is also relevant. We can consider $k$ rectangular blocks of 
length $l N $, where $l$ is a positive integer of order $1$ 
as $N$ goes to infinity. These 
Young diagrams also give contributions of the type $ e^{-A N}$, 
as can be seen from the formula 
The Casimir of $ ( k , l , R_1 , R_2 ) $ can be written as 
\eqn\sucas{\eqalign{  
& { C_{2}^{SU} ( k , l , R_1 , R_2 ) \over N }  = 
N kl ( l+1) +  n(R_1) +  n(R_2) - k^2l( l +1  ) +  2l n(R_2) \cr 
&  +  { 2 kl ~ ( ~ n(R_1) + n(R_2) ~ ) \over N }  +
 { \tilde C ( R_1) + \tilde C ( R_2) \over N  }
-  { n^2 ( R_1 ) \over N^2 } - { n^2 (R_2) \over N^2 } 
 \cr 
}}
Here $ \tilde C ( R ) = \sum_{i} r_i ( R ) ~  (~  r_i ( R )  -2 i +1 ) $
following the notation of \grotayi. 
It is clear from this formula that for Young Diagrams 
with $l$ being any number of order $1$ as $N$ goes to infinity, 
we get contributions  to the partition function \exctans , 
which go like finite powers of 
 $e^{-A N}$.  In section 2 we will consider a generalization 
 where $ ( k , l , R_1, R_2 )$ is replaced by Young Diagrams 
characterized by two vectors of integers $ \vec k , \vec l $
with $p$ components as well as small Young Diagrams 
$ R_1 , R_2 \cdots R_{p+1}$. 
In section 3, we consider composite Young Diagrams 
built from these generalized chiral Young Diagrams, 
together with their anti-chiral counterparts. 
We compute the Casimirs and their corresponding 
contribution to the partition function of two dimensional
 Yang Mills.  
In section 4,  we discuss the mixing between chiral and 
anti-chiral contributions  of these
composite Young Diagrams, and observe that the 
structure describing this mixing given in \vaf
for the $1/N$ expansion continues to hold for the $e^{-AN}$ terms. 

In section 5, we discuss the physical interpretation of 
the $e^{-AN}$ contributions. The plausible suggestion that they 
are related to D1-branes has been made in \lmr.   
This is discussed further and the extra parameter 
$l$  ( and more generally $ \vec l $ )  poses an interesting challenge
since it leads to $ e^{ -A N k l (l+1  )} $ which can be 
interpreted in terms of D-string wrappings only if the 
effective brane tension seen on the $T^2$ base space of 
the Yang Mills is proportional to $l(l +1)$.  One  
striking feature of the proposal in  \vaf\  is 
that the string dual for Yang Mills on $T^2$ involves 
a six-dimensional Calabi-Yau  geometry, which is a bundle 
$ {\cal{L}}_{m} \oplus {\cal{L}}_{-m}$ over $T^2$. More generally 
the Calabi-Yau would naturally sit in a ten-dimensional 
background.  Our attempts to explain the $l(l+1)$ rely on these 
extra dimensions of the geometry.

\newsec{  $e^{-AN}$ effects in $SU(N)$ 2dYM and large chiral Young Diagrams
 }

 The Young Diagrams of \lmr\ involve a block with 
 $k$ rows of length $N$. We mentioned above that
 we can also have $k$ rows of length $lN$ where $l$ is a 
positive integer of order $1$ as $ N \rightarrow \infty$. 
  There are is further simple generalization, where we have several 
 such blocks. 
 These Young Diagrams with several large blocks can 
 also have smaller, order $1$ Young diagrams attached to them.
 Figure 1 describes these Young Diagrams.  
 Suppose we have one such block with $k_1 $ long rows of length 
 $l_1N$. To the right of it we attach another Young Diagram 
 which has $k_2$ long rows of length $l_2N$. There is a restriction 
 $ k_1 > k_2 $. Further to the right we have  a block with $k_3 $ 
 long rows of length $l_3N$, and $ k_3 > k_2 $. And on up to $(k_p , l_p)$. 
 To this we can attach a small Young diagram ( with order $1$ boxes as 
 $ N $ goes to infinity )  to the bottom of the first $ ( k_1,  l_1)  $   
 block.  Another small Young diagram is added to the bottom of the second 
rectangular block, and so on up to the $p$'th block, and finally a small 
Young diagram can be attached to the right of the last block. 
These can be labelled $ R_1 , R_2  \cdots  R_{p+1} $. In Figure 1, we have 
the case $p=3$. The row lengths for the Young Diagram $R_1$ are all positive, 
since negative rows for it would imply negative row lengths 
for the large chiral Young Diagram, which are not allowed 
for $SU(N)$.  The row lengths for $R_2, \cdots R_{p+1} $ 
are allowed to be negative, subject to the usual restrictions 
$r_{i+1} \le r_i $. Negative row lengths imply that the large 
chiral Diagram has a few boxes chipped away from the blocks of 
sizes $ ( k_1, N l_1 ) \cdots (k_p , Nl_p ) $.  An example for $p=1$ is shown 
in Figure 3. Since the row lengths of $R_2 \cdots R_{p+1}$ are small 
compared to $N$, these negative row lengths only make small modifications 
to the large blocks.

The basic formula for the Casimir of $SU(N)$ is 
\eqn\bascas{ 
C_2 = N n + \sum_i r_i ( r_i - 2i +1 ) - n^2/N 
}   
where $r_i$ are row lengths and $n$ is the total number 
of boxes in the Young Diagram.



This large chiral Young diagrams
and their Casimirs can be described with the 
following notation. 
Define 
\eqn\hatkl{\eqalign{  
 \hat k_a &= k_a -  k_{a+1} \hbox{ for } a = 1  \cdots  p-1 \cr 
 \hat k_p &= k_{p} \cr 
 \hat l_a &= l_1 + l_2 + \cdots  + l_{a } \cr 
  n_0 &= \sum_{a=1}^{p} \hat k_a \hat l_a =
 \sum_{a=1}^{p}  k_a  l_a \cr
  n_1 &= \sum_{a=1}^{p+1} n ( R_{a} ) \cr   
}}

So we have described a  Young Diagram made out of 
large blocks described by $ ( \vec k , \vec l ) $ and 
small Young Diagrams  $ R_1 , R_2 \cdots R_{p+1}$. 
The row lengths for this large chiral Young Diagram 
are given by 
\eqn\chirlngths{\eqalign{ 
\hbox{For} ~~ i &= 1 ~ \cdots  ~ k_{p} \cr 
r_i ( R )    &=  N \hat{l}_p + r_i ( R_{p+1} )
\cr 
\hbox{For} ~~ i &= (k_{p}+1) ~ \cdots ~  k_{p-1 } \cr 
r_i ( R ) &= N \hat{l}_{p-1} + r_{i - k_p}    ( R_{p} ) \cr 
 & ~~~ \vdots \cr  
\hbox{For} ~~i &= (k_{2}+1) ~ \cdots  ~ k_{1} \cr 
r_i ( R )    &= N  \hat{l}_{1} + r_{i-k_{2}} (R_2)    \cr 
\hbox{For} ~~i &= ( k_{1}+1 ) ~ \cdots  ~  N \cr 
r_i ( R )    &= r_{i-k_1} ( R_1 ) \cr }   }
A more compact way to write this 
\eqn\chirlngthsi{\eqalign{
&  \hbox{For} ~~~ a = p ~ \cdots ~ 1 ~~~~~~ 
 \hbox{For} ~~ i = ( k_{a +1 } +1 ) ~  \cdots  ~  k_{ a } \cr 
& \qquad \qquad \qquad  
  r_{i}(R) = N {\hat l}_{a} + r_{i - k_{a+1} } ( R_{a+1})     \cr 
& \qquad \hbox{For} ~~i = ( k_{1}+1 ) ~ \cdots  ~  N \cr 
& \qquad \qquad \qquad  
r_i ( R )    = r_{i-k_1} ( R_1 ) \cr }   }
Note that, since $R_1$ is a small Young diagram with number 
of boxes small compared to $N$, the $r_i(R)$ given 
in the last line of \chirlngths\chirlngthsi\ 
 vanish well before $i$ reaches $N$.

Now the Casimir normalized by $N$ is described by 
\eqn\casnorm{\eqalign{  
{ C_2 \over N } & = N( ~  \sum_{a=1}^{p} \hat k_a \hat l_a ( \hat l_a + 1 )~  )
     \cr 
 &  + \bigl (  n_1 - n_0^2 - \sum_{a=1}^{p}  l_a  k_a^2+ 
2 \sum_{a=1}^p \hat l_a n ( R_{a+1} ) 
        ~~ \bigr ) \cr 
  & + { 1 \over N } ~ \bigl (  ~ n_1 - 2  n_0  n_1 + \sum_{ a =1 }^{p+1} 
 \sum_{i=1}^{\hat k_{a-1} } r_i^2 ( R_a )  - 2 \sum_{a=1}^{p} k_a n (R_a) 
   - \sum_{a=1}^{p} \sum_{i=1}^{ \hat k_{a-1}} 2 i r_i ( R_a ) ~  \bigr ) \cr 
 & -  { 1 \over N^2 } \sum_{a=1}^{p+1 } ( n^2 ( R_a ) )  \cr 
& = N c_0 + c_1 + { c_2 \over N } + { c_3 \over N^2 } \cr 
}}

In the last line we have defined $ c_0 \cdots  c_3 $
 which are to be read off from the previous equation. 
The term  $ \sum_{a=1}^{p}  l_a  k_a^2 $ can also be written 
as $ \sum_{a=1}^{p}  \hat l_a   ( k_a^2 - k_{a+1}^2 )  $ where 
$ k_{p+1} = 0 $. 

 Redefining $ { g_{YM}^2 N A \over 2 } \rightarrow A $, 
 the exact YM2 partition function \exctans\ can be written as  
\eqn\exct{ 
Z = \sum_{R } e^{- A C_2/N } 
}
Using the above decomposition of the Casimir for 
choices $ k_1 ~  \cdots ~ k_p $ and $ l_1~ \cdots ~ l_p $
we can write 
\eqn\zexp{ 
Z = Z_{pert} + \sum_{p=1}^{\infty} \sum_{k_1 \cdots k_p = 1 }^{\infty }
  \sum_{l_1 \cdots l_p = 1 }^{\infty }  \sum_{R_1 \cdots R_{p+1} }
  e^{ -A N c_0 -A c_1 - A c_2/N - A c_3 / N^2 } }
where the part perturbative in the $ { 1 \over N } $ expansion 
is the chiral Gross-Taylor expansion.

In the above we have built a ``large chiral Young Diagram'' 
with the data 
$$ [ k_1 , k_2 \cdots  k_p ,  l_1 , l_2 \cdots l_p ; R_1 \cdots R_{p+1} ]$$ 
where $k_1 \cdots  k_p $ and $l_1 \cdots  l_p $ are positive integers 
and $ R_1 \cdots  R_{p+1} $ are Young Diagrams. 
To be brief, we  denote them as $ [ \vec k , \vec l ; \vec R ] $.  
These Young diagrams have complex  conjugates. The composite 
Young diagrams built by putting together diagrams of the above 
type with conjugates can be denoted  as 
$$ [  k_1 , k_2 .. k_p ,  l_1 , l_2 ...l_p ; R_1 \cdots R_{p+1} ~ | ~  
  \bar k_1 , .. \bar k_q ,  \bar l_1 , ... \bar l_q  ;   S_1 \cdots
    S_{q+1} ]$$
More briefly we can denote them by 
$ [ \vec k ,  \vec l ;  \vec R ~|~  \vec  {\bar k} , \vec { \bar l }  ;
 \vec {  S }  
  ] $.     
We then need to compute the Casimirs of these non-chiral 
composites. We will do this in the next section.

\newsec{ $e^{-AN}$ effects : The non-chiral diagrams }




We consider the non-chiral Young Diagrams obtained 
by fusing a chiral Young Diagram and an anti-chiral Young Diagram. 
The chiral Young Diagram is specified by 
a pair of $p$-vectors of  integers $ \vec k , \vec l $
which specify rectangles of size $ k_1l_1N , k_2 l_2 N , \cdots $ 
and a set of $ p+1$ Young diagrams $ R_a $  of size order  $1$ as 
$ N $ goes to infinity. The $k$'s are ordered $ k_1 > k_2 > \cdots > k_p $.  
 The antichiral diagram is
specified by $ \bar k_1 \cdots \bar k_{q} $,  $ \bar l_1 \cdots \bar l_q $
and Young Diagrams $ S_1 \cdots S_{q+1} $. The $\bar k$'s are ordered 
$ \bar k_1 > \bar k_2 > \cdots > \bar k_q $. Figure 2 shows a non-chiral 
Young Diagram for the case $p=3, q = 3 $. Generalizing the possibility 
of negative row lengths  in the chiral case,
 row-lengths for $ S_2 , \cdots S_{q+1}$ can be negative. 

The row lengths $r_i$ of the composite Young Diagram $ R$  
are as follows : 
\eqn\rowlngths{\eqalign{ 
\hbox{For} ~~ i &= 1 ~ \cdots  ~ k_{p} \cr 
r_i ( R )    &= N (  \widehat   {\bar l_q}  + \hat  l_p )
  + r_1 ( S_{q+1} ) + r_i ( R_{p+1} )  
\cr 
\hbox{For} ~~ i &= (k_{p}+1) ~ \cdots ~  k_{p-1 } \cr 
r_i ( R ) &= N (  \widehat {  \bar l_q  } + \hat  l_{p-1}  )
 + r_1 ( S_{q+1} ) + r_{i-k_{p} }  ( R_{p} ) \cr 
 & ~~~ \vdots \cr  
\hbox{For} ~~i &= (k_{2}+1) ~ \cdots  ~ k_{1} \cr 
r_i ( R )    &= N  ( \widehat {  \bar l_q  } +  l_1  ) + r_1 ( S_{q+1} ) 
            + r_{i-k_2}  ( R_{2} )  \cr 
\hbox{For} ~~i &= ( k_{1}+1 ) ~ \cdots  ~  N -  \bar k_{1} \cr 
r_i ( R )    &= N   \widehat {  \bar l_q  }  + r_1 ( S_{q+1} )  
+ r_{i-k_1}  ( R_{1} )  - r_{N - \bar k_1 - i + 1 } ( S_1 ) \cr 
\hbox{For} ~~i &= ( N - \bar k_1 +1 ) ~  \cdots ~ (  N -  \bar k_{2} )  \cr 
r_i ( R ) &= N  ( \widehat {  \bar l_q  } -  \widehat { \bar l_{1} }  ) + 
r_1 ( S_{q+1} )   - r_{N - \bar k_2 - i + 1 } ( S_2 ) \cr 
& ~~~~ \vdots \cr  
\hbox{For} ~~i &= ( N - \bar k_{q} +1 ) ~   \cdots  ~  N  \cr
r_i ( R ) &=  r_1 ( S_{q+1} )   - r_{N  - i + 1 } ( S_{q+1} ) \cr 
}}
A more compact way to write this 
\eqn\rwlngths{\eqalign{
&  \hbox{For} ~~~ a = p ~ \cdots ~ 1 ~~~~~~ 
 \hbox{For} ~~ i = ( k_{a +1 } +1 ) ~  \cdots  ~  k_{ a } \cr 
& \qquad \qquad \qquad  
  r_{i}(R) = N  (  {\hat {  \bar  l}}_q   + \hat {l}_a   ) +
 r_1 ( S_{q+1} ) + 
                     r_{i - k_{a+1}}  ( R_{a+1} ) \cr 
& \qquad  \hbox{For} ~~ i = ( k_{1 } + 1)~   \cdots   ~ ( N - \bar k_{ 1 }) 
 \cr
 & \qquad \qquad \qquad 
 r_{i}(R) = N  \widehat {  \bar l_q  }+ r_1 ( S_{q+1} ) +
  r_{i-k_1}  ( R_{1} )  - r_{N - \bar k_1 - i + 1 } ( S_1 ) \cr
 &  \hbox{For} ~~~a = 1~  \cdots ~ q  ~~~~~~ 
 \hbox{For} ~~ i = (N - \bar k_{a}+1 )  ~  \cdots ~   
(N - \bar k_{ a +1 })  \cr 
 & \qquad \qquad \qquad 
r_{i}(R) = N  ( \widehat {  \bar l_q  }    - \widehat {  \bar l_a  } )
 + r_1 ( S_{q+1} ) - 
r_{N - \bar k_{a+1} - i + 1 } ( S_{a+1} ) \cr
}}
Although we are only using $p$-pairs of integers for the chiral Young Diagram 
$ k_1 \cdots k_p ; l_1 \cdots l_p $  and $q$ pairs $ \bar k_1 \cdots \bar k_q ;
 \bar l_1 \cdots \bar l_q $ for the anti-chiral Young Diagram,
it is useful to define  $ k_{p+1}= \bar k_{q+1} =0 $ in \rwlngths. 

Using these values for the row lengths we can compute 
the Casimir using \bascas\ and we find 
\eqn\casform{\eqalign{  
&{ C_2 \over N } =  N ( ~~ \sum_{a=1}^{p} \hat {k}_a 
\hat {l}_a( \hat {l}_a +1 ) +
 \sum_{a=1}^{q} \hat {\bar k}_a \hat {\bar l}_a 
( \hat{ \bar l}_a +1 ) ~~ ) \cr 
 & - ( n_0 - \bar n_0 )^2 
- \sum_{a=1}^{q} { \bar l }_a {\bar k}_a^2 - 
\sum_{a=1}^{p}  {  l }_a  k_a^2  \cr 
&  + \sum_{a =1 }^{p+1} n(R_a) +\sum_{a=1}^{q+1} n ( S_{a} )  
  + 2 \sum_{a=1}^{p} {\hat l }_{a} n ( R_{a+1} ) 
 + 2 \sum_{a=1}^{q} \hat {\bar l}_{a} n ( S_{a+1} )  \cr & 
\qquad \qquad \qquad \qquad \cr
& + { 1 \over N } ( ~~\sum_{a=1}^{p+1} n ( R_a ) + \sum_{a=1}^{q+1} n ( S_a )
 ~~  )
  \cr 
& - { 2 \over N } ( ~~ \sum_{a=1}^{p } k_a  n ( R_a ) + \sum_{a=1}^{q}
 {\bar k}_a n ( S_a ) ~~ ) \cr
& - { 2 \over N } ( n_0 - \bar n_0 ) (~~  \sum_{a =1 }^{p+1} n(R_a) +
 \sum_{a=1}^{q+1} n ( S_{a} ) ~~ ) \cr 
& + { 1 \over N } \sum_{a=1}^{p+1} \sum_{i} r_i^2 ( R_a ) + { 1 \over N }
 \sum_{a=1}^{q+1} \sum_{i} r_i^2 ( S_a )  - { 1 \over N }
\sum_{a=1}^{p+1} \sum_{i} 2 i r_i ( R_a ) - { 1 \over N }
\sum_{a=1}^{q+1} \sum_{i} 2 i r_i ( S_a ) \cr 
& - { 1 \over N^2 }  ( ~~ \sum_{a =1 }^{p+1} n(R_a) - 
 \sum_{a=1}^{q+1} n ( S_{a} ) ~~  )^2 
\cr  }}
We have defined anti-chiral analogs of \hatkl\ 
\eqn\barhatkl{\eqalign{  
& \hat { \bar l }_{ a } = \bar l_1 + \bar l_2 + \cdots + \bar l_a \cr  
& \hat { \bar k }_{ a } = \bar k_a - \bar  k_{a+1} \cr 
& \bar n_0  = \sum_{a=1}^{q} {\bar k}_a  {\bar l}_a =
 \sum_{a=1}^{q} {\hat {\bar k}}_a  \hat { {\bar l}}_a \cr 
}}

Note that this formula is symmetric under exchange of 
chiral and anti-chiral, which exchanges $ p \leftrightarrow q $ , 
$R_a \leftrightarrow S_a $. It also agrees with the chiral 
formula \casform\ when we set to zero the anti-chiral variables, 
 $ r_i (S_a) , n ( S_a ) , \bar k_a , \bar l_a , \bar n_0 $.

From \casform\ we can read off 
$ { C_2 \over  N } = N d_0 + d_1 + { d_2 \over N } + { d_3 \over N^2 } $ 
and we get an expansion for the parition function \exctans\   of the form. 
\eqn\expnchir{\eqalign{  
& Z = Z_{nchir}^{pert } + 
 \sum_{p=1}^{\infty} \sum_{k_1 \cdots k_p = 1 }^{\infty }
  \sum_{l_1 \cdots l_p = 1 }^{\infty }  \sum_{R_1 \cdots R_{p+1} }
  e^{ -A N c_0 -A c_1 - A c_2/N - A c_3 / N^2 } \cr 
& + \sum_{q=1}^{\infty} \sum_{\bar k_1 \cdots \bar k_q = 1 }^{\infty }
  \sum_{\bar l_1 \cdots \bar l_q = 1 }^{\infty }  \sum_{S_1 \cdots S_{q+1} }
  e^{ -A N \bar c_0 -A \bar c_1 - A \bar c_2/N - A \bar c_3 / N^2 } \cr 
& + \sum_{p=1}^{\infty} \sum_{q=1}^{\infty} 
  \sum_{k_1 \cdots k_p = 1 }^{\infty }
  \sum_{l_1 \cdots l_p = 1 }^{\infty }
  \sum_{\bar k_1 \cdots \bar k_q = 1 }^{\infty }
\sum_{\bar l_1 \cdots \bar l_q = 1 }^{\infty }
 \sum_{R_1 \cdots R_{p+1} }
  \sum_{S_1 \cdots S_{p+1} }  e^{ -A N d_0 -A d_1 - A d_2/N - A d_3 / N^2 }
}}
We have separated terms which are a series 
in $ 1/N$, the terms which involve $e^{-AN} $ and are purely chiral 
 ( written out before in \casnorm ) , the terms which   involve $e^{-AN} $ and 
 are purely anti-chiral, and finally terms involving $e^{-AN}$  
which are composites of chiral-antichiral. The constants 
$c_0 \cdots c_3 $ have been defined before and $ \bar c_0 \cdots \bar c_3 $ 
are obtained by changing $R_a$ to $ S_a $ , $ p $ to $q$ , $  ( k,l )  $ 
to $ ( \bar k , \bar l ) $.   The sums over $k, \bar k $ are constrained 
by $ k_1 > k_2 > ... > k_p $ and $ \bar k_1 >  \bar k_2 > \cdots \bar k_q $  
as is clear from Figures 1 and 2.  We do not have a proof that the 
exponential contributions in \expnchir\ are complete. In fact one can write 
other types of diagrams which have $C_2 \sim N^2 $, e.g a Diagram 
 having row lengths starting from $ \sqrt {N} $ and dropping in steps of 
$1$ down to zero, or a hook-shaped Young Diagram with first row 
of length $N/2$ and first column of length $N/2$. 
 An important question is to find a complete description 
of the exponential corrections. It is also possible that   the subset of 
 terms picked out in \expnchir\ have a physical meaning in terms of D-branes 
 in the equivalent topological string theory. D-branes will be 
discussed in further in section 5. 
Even if we have a complete description of the 
$e^{-AN}$ terms, 
it is not clear that this would capture the full structure 
of the finite $N$ partition function.
 For example we can certainly choose certain
classes of Young Diagrams which contribute terms of the type 
$e^{-AN^2}$. It is  worth noting that other scaling limits 
of 2dYM can be studied \gss. 
 Whether the  $1/N$ expansion and the exponential terms of the type
 $ e^{-AN}$ can allow us to reconstruct the finite $N$ answer
 could be studied using techniques of exponential 
 asymptotics \expoas.

\newsec{ Holomorphic/Antiholomorphic  mixing 
in $SU(N)$ and $U(N)$ partition functions } 

Define $B$ as the number of boxes in the chiral Young Diagram 
and $ \bar B$ as the number of boxes in the antichiral Young Diagram. 
\eqn\boxs{\eqalign{  
B &= b_1 + b_0 \cr 
\bar B &= \bar b_1 + \bar b_0  \cr 
b_1 &= n_1 =  \sum_{a=1}^{p+1} n(R_a) \cr 
b_0 &= N n_0 = N \sum_{a=1}^{p} k_a l_a \cr 
\bar b_1 &= \bar n_1 = \sum_{a=1}^{q+1} n(\bar R_a) \cr 
\bar b_0 &=  N \bar n_0 = N \sum_{a=1}^{q} { \bar k_a}  { \bar l_a } \cr 
}} 

An important observation about the sum in \casform\ 
is that the mixed terms involving products 
of chiral and anti-chiral can be expressed 
entirely in terms of $ ( B - \bar B )^2 $. 
Indeed all such mixed terms in \casform\ are included in : 
\eqn\mxd{ 
 - ( n_0 -  \bar n_0 )^2 - { 2 \over N } ( n_0 - \bar n_0 ) 
( \sum_{a=1}^{p+1} n(R_a) - \sum_{a=1}^{q+1} n(S_a) )
-{ 1 \over N^2 } ( b_1 - \bar b_1 )^2   
} 
which is exactly equal to $  -  ( B - \bar B )^2    $ 
when we use the fact
\eqn\bnrels{\eqalign{  
  B &= b_0 + b_1 = N n_0 + \sum_{a} n(R_a) \cr 
 \bar B &= \bar b_0 + \bar b_1 =  N \bar n_0 + \sum_a n(S_a) \cr 
}}

Defining \foot{  Note that the partition function of 2dYM 
depends on $ g_{YM}^2  A , N $. In previous 
sections  we had redefined  
$ { g_{YM}^2 N A \over 2 }  \rightarrow A $. Here 
we are setting $A=1$ for comparison to \vaf .  }
$$ Z_+ = \sum_{R_+} exp ( - { 1 \over 2  } g_{YM}^2 C_2(R_+ ) + i \theta B )$$ 
we find that it can be written as 
\eqn\chp{ 
Z_+ = \sum_{R_+ } exp ( -{1\over 2 }g_{YM}^2 \kappa ( R_+ ) - t |R_+| )} 
where  $ t=  { N g_{YM}^2 \over 2 } - i \theta $ and 
we separated  $ C_2 (R_+ ) = \kappa (R_+ ) + N B $.     
This can be viewed 
as a definition of $ \kappa (R_+ ) $, and the explicit formula 
for it can be read off from \casnorm 
\eqn\kapform{\eqalign{ 
&\kappa( R_+ ) = N  \sum_{a=1}^{p} \hat k_a \hat l_a^2  \cr 
 & -n_0^2   + 2 \sum_{a=1}^p \hat l_a n ( R_{a+1} ) 
     - \sum_{a=1}^{p}  l_a  k_a^2   \cr 
  & + { 1 \over N } \bigl ( ~   n_1 - 2  n_0  n_1 + \sum_{ a =1 }^{p+1} 
 \sum_{i=1}^{\hat k_{a-1} } r_i^2 ( R_a )  - 2 \sum_{a=1}^{p} k_a n (R_a) 
   - \sum_{a=1}^{p} \sum_{i=1}^{ \hat k_{a-1}} 2 i r_i ( R_a ) ~ \bigr )  \cr 
 & -  { 1 \over N^2 } \sum_{a=1}^{p+1 } ( n ( R_a ) )^2 ~ .  \cr 
}}
When we neglect the exponential corrections 
and keep only the $1/N $ expansion, $ k_a = l_a = n_0 =  0 $
and the only Young Diagram appearing is $ R_1$. Then 
the formula for $ \kappa $    reduces, as expected, to 
$ \sum_{i} r_i(R_1)  ( r_i(R_1)  +1 - 2i ) $. 

 As we have observed above, 
all the mixed chiral-antichiral terms in \casform\ come from 
$ ( B- \bar B )^2 $. As a result, we can write
\eqn\nchir{ 
 Z_{SU(N)} = exp[ { g_{YM}^2\over 2N } 
( \partial_t - \partial_{\bar t } )^2  ] Z_{+} (t) Z_- (\bar t)
}

 The number of boxes in the chiral  $SU(N)$ Young Diagram 
 is $B $ , and the number of boxes in $ \bar  S  $ 
 ( the conjugate of the  anti-chiral Young Diagram $S$ )  is 
$ N r_1 ( S ) - \bar B   $. Using the description 
 of the Young Diagrams given before,
 $ r_1 (S) = r_1 ( S_{q+1} ) + N \sum_{a=1}^{q} \bar l_a$. 
 To the $SU(N) $ \YD\ we append a number $r$ of columns of length $N$ 
 to the left.  
 The $U(1) $ charge is $ q = N r + N r_1(S) + B - \bar B = N l + B - \bar B $, 
 using the definition $l = r + r_1(S)$.  

 To get the $ U(N) $ partition function, we use
 $C_2^{U(N)} = C_2^{SU(N)} + q^2/N$ and \exctans\  
 to write 
\eqn\unpart{\eqalign{ 
Z_{U(N)} &=   exp( - { g_{YM}^2 q^2 \over 2 N } ) Z_{SU(N)} \cr 
&=  exp( -  {g_{YM}^2 \over 2N} ( Nl + B  - \bar B )^2 ) Z_{SU(N)} \cr
}}

 As in the discussion of \vaf, the mixed term 
$  exp( -  {g_{YM}^2 \over 2N} ( B  - \bar B )^2 )$
 cancels with the mixing term in the $SU(N)$ partition 
function. 

 Then the same steps lead to 
\eqn\unform{ 
\tilde Z_{U(N)} = \sum_{l= -\infty}^{\infty} 
\tilde Z_+( t + l g_{YM}^2 ) \tilde Z_- ( \bar t - l g_{YM}^2 ) 
}

Where $\tilde Z_{+} (t )  =  Z_{+} (t) e^{ -{ t^3 \over 6 g_{YM}^4 } +
 { t \over 24 } }  $ and $ \tilde Z_{U(N)} = 
e^{ { - ( t^3 + \bar t^3 ) \over 6 g_{YM}^2 }
 + { t + \bar t \over 24 } } Z_{U(N)} $. 
This is the same form as in \vaf, but the 
$Z_+$ and $\kappa(R_+) $ entering this formula 
 are different and are given in \chp\ and \kapform. 

\newsec{ D-branes and $e^{-AN}$  }

\subsec{ 5-brane and Myers effect ?  }

The $ l=1$ contributions to the partition function 
containing factors  of $ e^{ - AkN   } $  were interpreted 
in terms of D-strings wrapped on the torus with winding number 
$k$.  Now that we have found  the extra parameter $l$ 
 in the space of Young diagrams contributing $ e^{-AN } $ 
 effects and  are getting $ e^{ - A k l ( l +1 ) N } $, we 
 can ask what is the interpretation of the factor 
 $ k l ( l +1 ) $. Note that the  discussion in \vaf\  is in a IIA 
 superstring theory context. For convenience the current discussion 
is in a IIB context. But  the current 
discussion could be translated to a IIA set-up by considering 
$2$-branes  instead of $1$-branes ( as well $4$ and 
 $6$-branes rather than $3$ and  $5$-branes 
in the following )  and it may also 
be interesting to study the effects of T-duality relating IIA and IIB. 

One possibility is that the factor of $ l ( l+1) N = l  ( l+1) / g_s $ 
 should be interpretd as a modified tension of D-strings. 
 In the proposal of \vaf\  for the string theory of 
 2dYM there are elementary strings on a CY which is a 4D space fibered 
 over the $T^2$ of the 2dYM theory. 
  We can get D-strings 
 with modified tension, if the D-strings are actually extended 
 objects in the extra directions. For example they could be 5-branes 
which wrap  the $T^2$ as D-strings do,  but extend 
 into the extra directions. What kind of transverse geometry 
 can explain the form $ l(l+1)$ ? 

 The form $l(l+1)$  can be understood if we make an assumption about 
 the geometry of these D-strings in the directions 
 transverse to the $2$-torus base. We assume that they are 
 not points in the transverse dimensions but rather 
 form a fuzzy geometry by the Myers dielectric effect 
 \myersdiel.  
 It is a familiar fact in fuzzy geometry 
 that when a D-string expands  into a 5-brane to 
 form a funnel with fuzzy 4-sphere cross-section, the number 
 of D-strings involved cannot be an arbitrary integer 
 but rather is constrained to be integers of the form 
 $ { (n+1) ( n +2 ) ( n+3 ) \over 6}  $ where $n$ is a positive  integer
\refs{ \nabint , \clt , \ram }.  
 If the transverse geometry is a fuzzy $ CP^2 $ 
\refs{ \nara , \grostei } 
then 
 the allowed numbers of D-strings are dimensions 
 of symmetric representations of $SU(3)$ which, for 
 a representation with $p$ boxes is $ { ( p +1) ( p+2  ) \over 2 }  $. 
 Identifying $ (p +1) $ with $l$ above, we have a  picture 
 where $l(l+1)$ D-strings bound to form a fuzzy $CP^2$ wrap the $T^2$ 
 with degree $k$.

 Given that the extra dimensions of the Calabi-Yau are 
just given by fibres of line bundles $ L_m \oplus L_{-m} $, 
 there is no obvious $CP^2$ in the picture. The topological 
string of the CY is of course embedded in the physical 
string theory, which has an extra $R^4$ transverse to the 
CY. The $CP^2$ might involve directions internal to the CY, 
 as well as directions transverse to the CY. It  might also  require 
 taking a near horizon geometry of branes wrapping  
a $4$-cycle on the CY, since the gauge theory on such 
 branes is what gives rise to the 2dYM in the picture of 
 \vaf.

\subsec{ Other possibilities : 3-brane ?  } 

 Another possibility is that the transverse geometry  
 of the branes rsponsible for the $e^{-AN}$ effects is
  two-dimensional. So the D-strings are secretly 
 a D3-branes. Two of their  worldvolume directions wrap around the 
 $T^2$ base of the Calabi-Yau and the remaining two directions 
 wrap directions internal to the fibre over the $T^2$. 
 If the winding number in the internal directions 
 is restricted to be of the form $ l( l+1)$,  
 then from the point of view of the $T^2$ base, the 
 $D3$-brane looks like a $D1$  
 string with effective tension proportional to $l(l+1)$. 
 It is not clear what mechanism would restrict 
 the winding numbers to be of the above form. 
 It is possible that we have to use the $K$-theory 
 approach to D-brane charges \refs{\Witten , \moore } 
 in order to understand the restricted form of the winding numbers.

\subsec{ Stringy excitations of D-brane systems } 

 The detailed 
structure of \casnorm\ and \casform\
should contain enough information to distinguish 
 between the possibilities above. 
 A given term going like  $e^{-AN}$ is multiplied by a series $ P(1/N ) $ 
 which should be accounted by strings stretching between 
the D-branes. A lot of the terms in \expnchir\ are directly 
related to the quadratic Casimirs of the Young Diagrams 
$ R_1 ... R_{p+1} $ and $ S_1 ... S_{q+1}$. In the case where
we restrict attention to perturbative ( $ 1/N$ ) terms, 
there is only $R_1$, $S_1$ which have been interpreted in terms 
of closed elementary strings \refs{ \grotayi , \grotayii }. 
 The full set $R_1 \cdots R_{p+1}$ should be related to 
closed as well as open elementary strings, stretching between 
different   $D$-branes, where the $D$-branes are labelled 
by $ ( \hat k_a , \hat l_a ) $. Since $R_1$ does not appear 
in second line of \casnorm\ it is distinguished from the $R_2 \cdots R_{p+1}$. 
We have also noted that the latter can have negative row lengths. 
The negative row lengths correspond to a left-right and up-down 
reflected Young Diagram. The positive part of $R_2 \cdots R_{p+1}$ 
can be associated with open strings  starting and ending on the 
$p$ D-branes and mapping  holomorphically to the target. The 
negative parts can be associated to similar open strings mapping 
anti-holomorphically to the target. The presence of $p$ sectors of 
such open string fluctuations suggests the D-branes are in 
phase where $ U(p)$ is broken to $U(1)^p$. There may also be
a closed string description of the excitations around the D-brane. 

 The integers $ ( \bar k , \bar l ) $  characterizing the 
 anti-chiral parts  of the composite  Young Diagrams would correspond to 
anti-branes. 
$S_2 \cdots S_{q+1} $ would be related to stringy excitations  
of the $q$ anti-D-branes. The positive and negative parts  would map  
 to $T^2$ holomorphically and anti-holomorphically
respectively.  The precise nature of the D-branes 
 will  affect the form of the coupling between the open strings 
and the D-branes.  Hence the D5 versus
 D3 possibility could be resolved. We leave this for future work.

Another interesting avenue in connection with the spectrum of 
branes that might be relevant to an understanding of the 
exponential type terms in 2dYM involves fractional D-strings. 
While all the terms considered so far ( and in the bulk of the 
paper )  involve integral 
powers of $ e^{-AN}$, we can also get fractional powers. 
The large blocks we considered so far have lengths $lN$, 
where $l$'s are integers. If we allow these to be fractional, 
the terms of type $ e^{ -ANl(l+1) }$ will give fractional powers 
of  $ e^{-AN}$. For example if $l = { 1 \over m}$  with $m$ large 
compared to $1$ but small compared to $N$ we can get $ e^{-AN/m}$. 
These appear to be contributions from fractional D-strings. 
Fractional branes have been discussed before in 
\refs{ \maldasuss\brodie\ferr }

\newsec{ Conclusions  } 
 
 We have computed a class of non-perturbative 
 contributions to the partition function 
 for two-dimensional Yang Mills theory on a torus, 
 behaving like powers of $e^{-AN}$ multplied by series 
 in $ 1/N$. These contributions come from a class of 
 large Young Diagrams, generalizing the ones that were introduced 
 in this context in \lmr.  
 It will be interesting to investigate whether the 
 exponential terms we have described are in an appropriate sense 
 complete e.g whether they provide the complete description of 
 D-branes in the dual topological string theory. 
 The free fermion description of 2dYM may be useful 
 in exploring this question.

 We showed that structure of the mixing between 
 chiral and anti-chiral sectors,  for the $SU(N)$ and the 
 $U(N)$ partition functions,  is very similar to the one 
 described in \vaf\  for the $1/N$ expansion.  
 We discussed the geometrical interpretation of the 
 non-perturbative contributions in terms of D-branes. 
 This discussion is far from complete, but it suggests
 that a proper geometrical understanding requires 
 the use of more than the $T^2$ base of the two-dimensional 
 Yang-Mills and should involve the Calabi-Yau described 
 in \vaf\ or the full ten-dimensional background of a physical string 
 related to the topological string of the CY. One of the approaches 
 we discussed for explaining the detailed form of the exponential 
 contributions involves  D-strings blowing up into a D5-branes 
 having a transverse fuzzy $CP^2$ geometry. Further clarifying 
 the D-brane picture is a very interesting direction for the future. 

 It will be  interesting to explore the meaning of the 
 non-perturbative holomorphic-antiholomorphic 
 mixing in the light of connections 
 to black hole entropy \refs{ \vaf , \osv , \ddmp , \sen}.  
 Large Young diagrams have also appeared recently 
 in the dual gauge theory description of non-perturbative 
 objects in ADS/CFT \refs{ \bbns , \cjr ,\aabf ,  \kg , \ber , \llm }.
 Large horizontal Young diagrams in that context are related 
 to ADS-giant gravitons. Whether there is a clear connection between 
 the geometry  of large Young Diagrams in ADS/CFT and 
in two-dimensional Yang Mills  is an interesting question. 
 
\bigskip

\noindent{\bf Acknowledgements:}
 We would like to  thank for useful discussions
 Simon Mc Namara, Costis Papageorgakis, 
   Bill Spence, Richard Szabo, Gabriele Travaglini,  Dan Waldram.  
 SR is supported by a  PPARC Advanced Fellowship held at Queen Mary, 
and in part by the EC Marie Curie Research Training Network
MRTN-CT-2004-512194.  
A. Jevicki is supported by DOE grant DE FG02/19ER40688 (Task A).
R. de Mello Koch is supported by NRF grant number 
Gun 2047219. 

\bigskip 

\vfill\eject

\epsfxsize=5.1in
\epsfbox{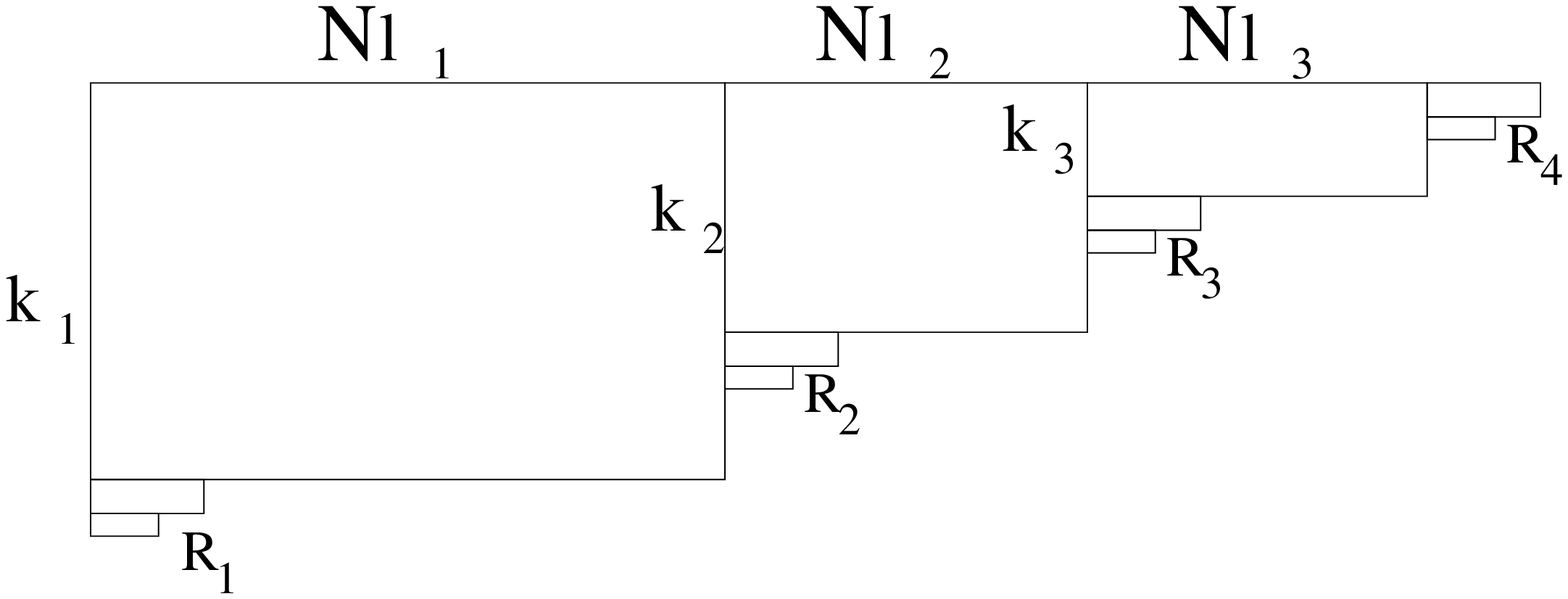}

{\bf  Figure 1 : Picture of a large chiral Young Diagram }

 The positive 
integers $k_i,l_i $ are or order $1$ as $N \rightarrow \infty $. 

\vfill\eject
\bigskip 
\epsfxsize=5.1in
\epsfbox{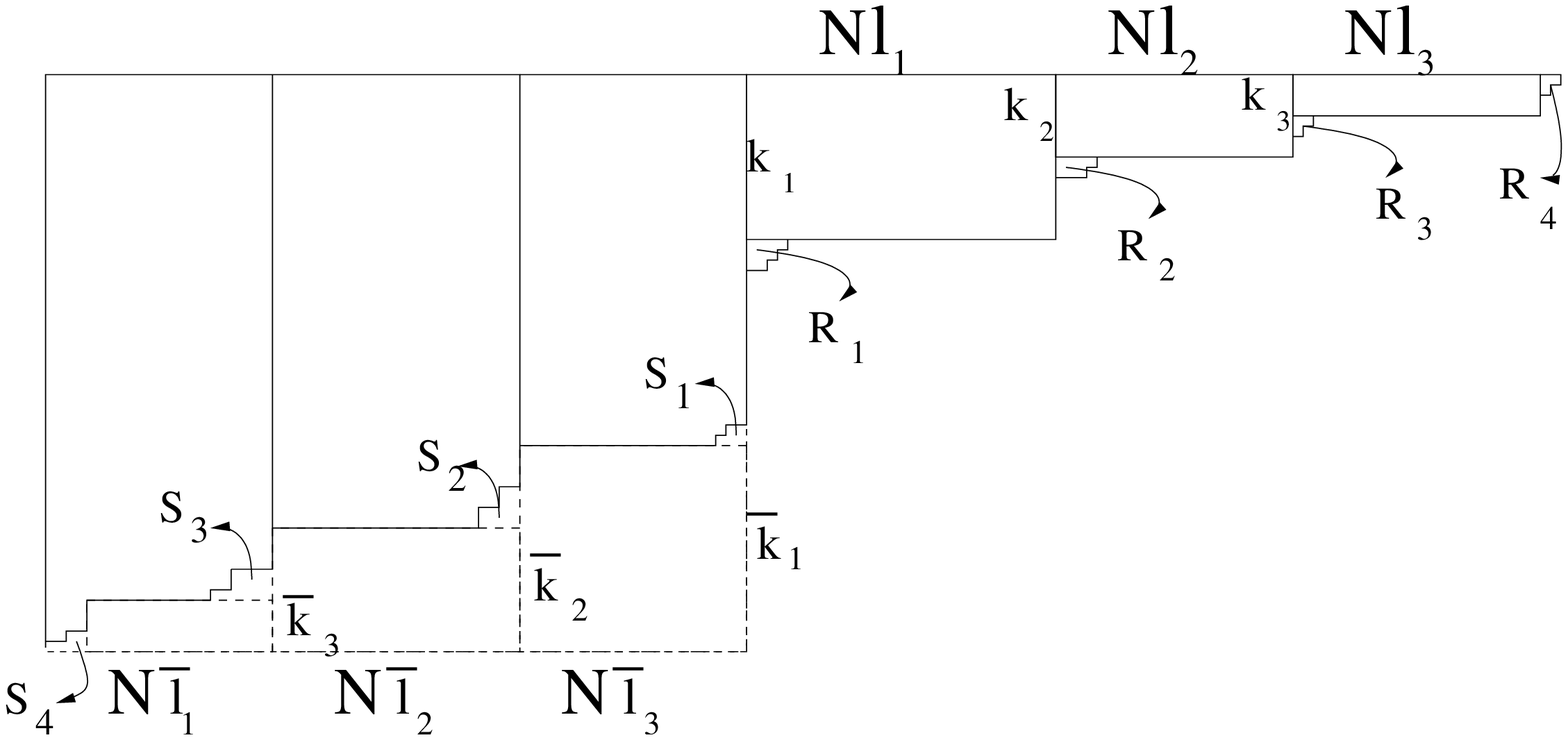}

{\bf  Figure 2 : Picture of a large non-chiral Young Diagram }

 The positive integers $k_i,l_i , \bar k_i , \bar l_i  $
 are or order $1$ as $N \rightarrow \infty $.

\vfill\eject
\bigskip 
\epsfxsize=5.1in
\epsfbox{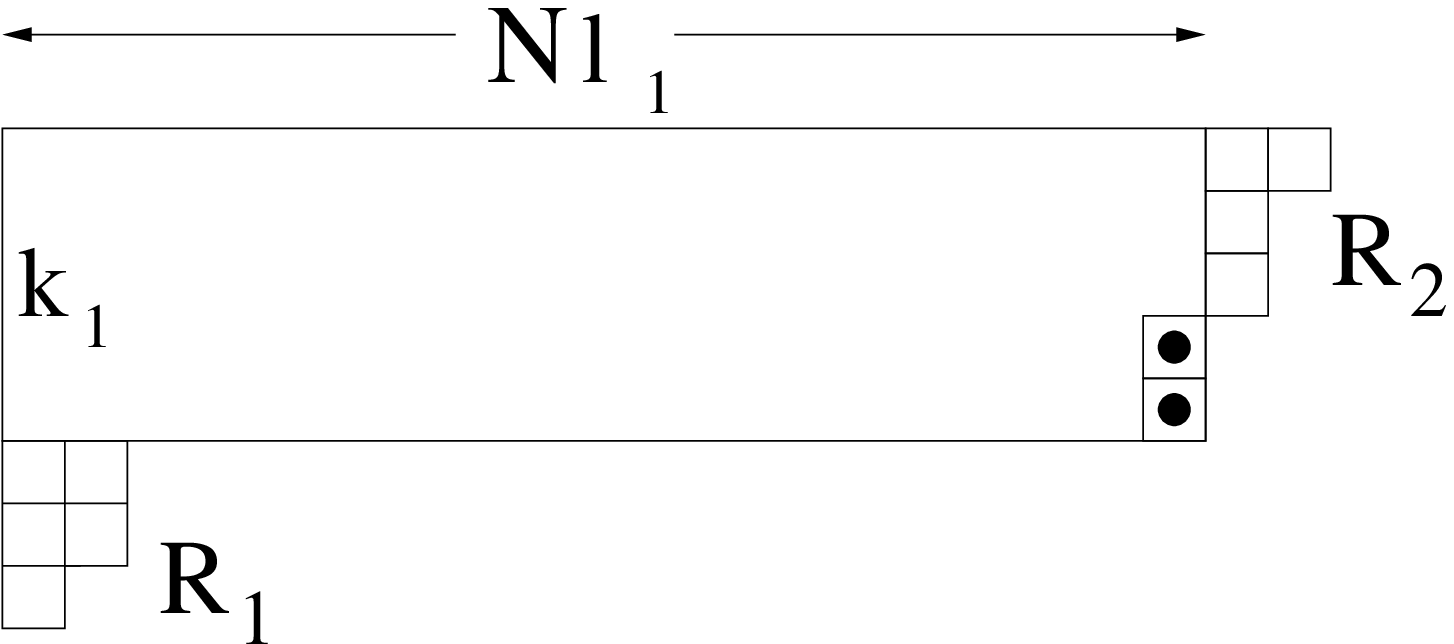}

{\bf  Figure 3 : Picture of large chiral Young  Diagram \hfill \break
  \hbox{                      } $~~~~~~~~~~~~~~~~~~$ 
  with negative row lengths for $R_2$.  }

 Example with $p=1$ showing $R_2$ having row lengths 
 $( 2, 1, 1, -1 , -1 ) $. The dotted boxes corresponding 
to the negative row entries are deleted from the large block 
of size $Nk_1l_1$.



\listrefs

\end